\documentclass[sigconf]{acmart}

\AtBeginDocument{%
  }

\copyrightyear{2026}
\acmYear{2026}
\setcopyright{cc}
\setcctype{by}
\acmConference[SIGIR '26]{Proceedings of the 49th International ACM SIGIR Conference on Research and Development in Information Retrieval}{July 20--24, 2026}{Melbourne, VIC, Australia}
\acmBooktitle{Proceedings of the 49th International ACM SIGIR Conference on Research and Development in Information Retrieval (SIGIR '26), July 20--24, 2026, Melbourne, VIC, Australia}
\acmDOI{10.1145/3805712.3809644}
\acmISBN{979-8-4007-2599-9/2026/07}

\newcommand{\system}{\textsc{DocQAC}{}}

\usepackage{multirow}
\usepackage{booktabs}
\usepackage[most]{tcolorbox}
\usepackage{arydshln} 
\usepackage[ruled, vlined, linesnumbered]{algorithm2e}
\usepackage[most]{tcolorbox}
\usepackage{xcolor}

\renewcommand{\acmConference}[4]{}

\begin{document}

\title{\system{}: Adaptive Trie-Guided Decoding for Effective In-Document Query Auto-Completion}

\author{Rahul Mehta}
\email{mehtarahul@microsoft.com}

\affiliation{%
  \institution{Microsoft Corporation
}
  \city{Hyderabad}
  \country{India}
}
\affiliation{
  \institution{Indian Institute of Technology}
  \city{Kharagpur}
  \country{India}
}

\author{Kavin R V}
\email{kavinrv13@gmail.com}
\affiliation{%
  \institution{Indian Institute of Technology Kharagpur}
  \city{Kharagpur}
  \country{India}
}

\author{Indrajit Pal}
\email{pal.indrajit99@gmail.com}
\affiliation{%
  \institution{Independent}
  \city{Bengaluru}
  \country{India}
}

\author{Tushar Abhishek}
\email{tabhishek@microsoft.com}
\affiliation{%
  \institution{Microsoft Corporation
}
  \city{Hyderabad}
  \country{India}
}

\author{Pawan Goyal}
\email{pawang.iitk@gmail.com}
\affiliation{%
  \institution{Indian Institute of Technology Kharagpur}
  \city{Kharagpur}
  \country{India}
}

\author{Manish Gupta}
\email{gmanish@microsoft.com}
\affiliation{%
  \institution{Microsoft Corporation
}
  \city{Hyderabad}
  \country{India}
}

\begin{abstract}
Query auto-completion (QAC) has been widely studied in the context of web search, yet remains underexplored for in-document search, which we term \system{}. \system{} aims to enhance search productivity within long documents by helping users craft faster, more precise queries, even for complex or hard-to-spell terms. Unlike traditional WebQAC systems, \system{} can leverage rich document context, having access not only to the partially typed user query and global historical queries, but also the content of the current document itself, and crucially, the document-specific history of user query interactions.
  
To address this setting, we propose a novel adaptive trie-guided decoding framework  that uses user query prefixes to softly steer language models toward high-quality completions. Our approach introduces an adaptive penalty mechanism with tunable hyperparameters, enabling a principled trade-off between model confidence and trie-based guidance. To efficiently incorporate document context, we explore retrieval-augmented generation (RAG) and lightweight contextual document signals such as titles, keyphrases, and summaries.

When applied to encoder–decoder models like T5 and BART, our trie-guided framework outperforms strong baselines and even surpasses much larger instruction-tuned models such as LLaMA-3 and Phi-3 in seen-query settings. This demonstrates its practicality for real-world \system{} system deployments, where efficiency and scalability are critical. We evaluate our method on a newly introduced \system{} benchmark derived from ORCAS, enriched with query–document pairs.
We make both the \system{} dataset\footnote{Dataset- \url{  https://bit.ly/3IGEkbH }\label{dataFN}} and code\footnote{Code-  \url{https://github.com/rahcode7/DocQAC}\label{codeDataFN}} publicly available.
\end{abstract}



\maketitle

\setlength{\tabcolsep}{2pt}
\begin{table}[!t]
    \centering
    \scriptsize
    \begin{tabular}{|p{0.2\columnwidth}|l|p{0.35\columnwidth}|p{0.35\columnwidth}|}
    \hline
    Document&Prefix&WebQAC output&\system{} output\\
        \hline
        \hline
     \url{https://en.wikipedia.org/wiki/Paris}&fr&freejobalert, freepik, free games, free fire max, from tv series, friends, friendship quotes, freecell, frank lampard, free job alert 2025&france capital, france tourism, france history, france landmarks, france culture, france travel guide, france famous cities, france eiffel tower, france paris attractions, france paris museums\\
         \hline
       \url{https://en.wikipedia.org/wiki/Brad\_Pitt}& a&amazon, adobe acrobat, anydesk, australia vs india, amazon prime, allahabad, aiden markram, american airlines, amitabh bachchan, alice in borderland&american actor, academy awards, angelina jolie, angelina jolie husband, aniston, alcoholism, academy award nominations, angelina jolie and brad pitt relationship, autobiography of brad pitt, a river runs through it\\
           \hline
        \url{https://history.house.gov/People/Office/Speakers-List/}& spea&speaker, speak no evil, speak, speaking, speaker cleaner, speak now, spear, speaker test, speaker cleaning, spearmint tea&speaker of the house, speaker of the house history, speaker of house of representatives, speaker history, speakers of the us house, speakers, speaker henry clay, speaker of the house current status, speaker of the house duties, speaker of the house responsibilities\\
            \hline
    \end{tabular}
    \caption{Examples of top few results from WebQAC versus \system{} systems.}
    \label{tab:samples}
\end{table}

\section{Introduction}

Query Auto Completion (QAC) is the first service with which search users interact, offering ranked query suggestions based on the partially typed user query (which we call a prefix). Traditionally, the most common approach to solving this task involves utilizing highly efficient trie-based data structures~\cite{hsu2013space} with techniques like Most Popular Completions (MPC)~\cite{bar2011context}. Recently, deep learning-based approaches that utilize sequence-to-sequence models trained on historical queries have been employed to generate high-quality completions~\cite{lee2021improving,mustar2020using,wang2020efficient,jiang2018rin}. We will refer to such QAC systems for Web search as WebQAC systems.

\subsection{Motivation}
While QAC has been extensively studied in the context of web search (WebQAC), where suggestions are driven by global popularity and historical query logs, there has been surprisingly no research dedicated to QAC for in-document search. We term this under-explored task as Document Query Auto-Completion (\system{}). \system{} systems can help users in (a) reducing search time by predicting and suggesting search terms, which in turn helps quickly find relevant information within long documents, and (b) improving search accuracy by minimizing typographical errors, especially for documents with rare and orthographically challenging terms. Overall, a \system{} system can enhance user productivity, ensure that users are using the correct terminology and phrases, and assist users who may not be familiar with the exact terms or keywords to use, making it easier for them to navigate and locate specific information without having to sift through long documents. 
The extra document context in \system{} systems brings an additional challenge. How do you best leverage this document context and associated metadata? How do you handle long documents?

\subsubsection{Difference between \system{} and  WebQAC}
A \system{} system differs from a WebQAC system in several fundamental ways.

\begin{itemize}
\item \textbf{Intent locality:} In WebQAC, user intent is often navigational (finding a specific website, e.g., ``youtube'') or broad information. In \system{}, user intent is tightly coupled to the current document and is exploratory in nature. As a result, globally frequent queries are often irrelevant, while rare or document-specific terms become critical for accurate completion.

\item \textbf{Vocabulary shift}: Documents frequently contain named entities, domain-specific phrases, and long-tail terminology that do not appear frequently in global query logs.

\item \textbf{Contextual Grounding}: The same prefix may require radically different completions depending on the document being viewed. Unlike WebQAC, which prioritizes global consensus, \system{} requires local contextual grounding: suggestions must be relevant to the specific content of the document at hand. This capability is critical for enhancing productivity in long-form document consumption, aiding users in navigating dense technical texts, and mitigating orthographic errors for rare, document-specific terms.
\end{itemize}

Table~\ref{tab:samples} illustrates this gap. For identical prefixes, WebQAC systems produce globally popular suggestions that are largely irrelevant to the document context, while a \system{} system must surface completions grounded in the document’s content and semantics.

\subsubsection{Use Cases of \system{}}
A \system{} system can appear in document interaction tools like Adobe Acrobat’s in-document search, text editors, IDEs, and enterprise document management systems. Here, users frequently write short prefixes to locate sections, entities, or phrases within a document. For example, when navigating long PDFs, reviewing technical documentation, or searching policy and legal documents. \system{} auto-completion in these scenarios improves efficiency by suggesting contextually relevant, document-specific terms during query formulation.



\subsection{Core Contributions}

\begin{figure*}[t]
    \centering
    \includegraphics[width=\textwidth]{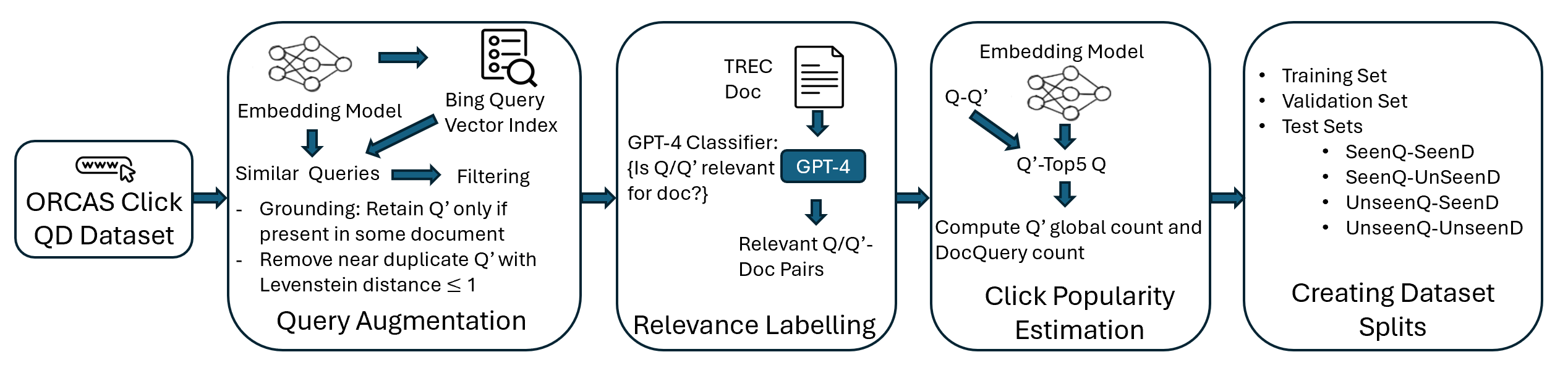}
    \caption{\system{} Dataset Construction Pipeline (detailed in Section~\ref{sec:dataset}). A document $D$ in ORCAS has clicked queries $Q$. Query Augmentation augments $D$ with non-clicked queries $Q'$ which are similar to $Q$. Relevance Labeling filters queries in $Q$ and $Q'$ that are irrelevant to $D$. Click Popularity Estimation estimates pseudo-counts for $Q'$ queries. Finally we create dataset splits.
    }
    \label{fig:datapipeline}
\end{figure*}

Overall, we make the following main contributions in this paper.
\begin{itemize}
    \item  \textbf{Formalization of the \textbf{\system{}} Task}: We formalize document specific Query Auto-Completion (\system{}) as a distinct problem paradigm, differentiating it from standard WebQAC by its strict faithfulness constraints and ``cold-start'' document challenges. To support this, we also release a dataset benchmark for the \system{} task.
     \item \textbf{Trie-Guided Inference Time Decoding Framework:} We develop a novel adaptive trie-guided decoding framework to softly bias encoder-decoder language model generation toward valid completions without adding new parameters. This method resolves the ``generative drift'' problem inherent in standard LLMs by dynamically pruning tokens that do not appear in the document's query log or body. 
      \item \textbf{Efficiency and Performance Gains}: We demonstrate that small, guided encoder-decoder models (e.g., T5-Small, BART-Base) significantly outperform unguided Large Language Models (LLaMA-3, Phi-3). This establishes a new state-of-the-art for efficient, latency-constrained query completion.
      \item   \textbf{Understanding effect of Context Representations}: We also perform detailed ablation studies to understand the role of document context, such as using document title and URL tokens, summaries or keyphrases extracted from documents.
\end{itemize}
To facilitate further research in this direction, we make the dataset\footref{dataFN} and code\footref{codeDataFN} publicly available.

\section{Related Work}
\label{sec:relatedWork}
\subsection{Autocompletion Methods}
\subsubsection{Trie-Based Methods}
     Given a prefix, the MPC model, proposed in~\cite{bar2011context}, extracts a limited number ($k$) of completions from a character-trie structure (also referred to as the main trie) built using a corpus of past queries. 
\subsubsection{Generative Methods}
    QueryBlazer~\cite{kang2021queryblazer} is a fully generative, low-latency query auto-completion method that is capable of leveraging both previously encountered queries and generating completions for new, unseen queries. 
    At inference time, the user’s query prefix is input into the subword encoder as a sequence of characters. The encoder then produces all potential top-k subword sequences that can be generated from the partial input provided.
    
    Deep learning methods like Hierarchical RNN Encoder-decoder~\cite{song2017hierarchical} with pointer generator~\cite{dehghani2017learning}, GRUs with user and time representations~\cite{fiorini2018personalized} and Transformer-based hierarchical encoder~\cite{yin2020learning} have also been studied. While showing suggestions it is important to not show defective suggestions and prefixes. To avoid defects, researchers have used LSTMs for inappropriate query suggestion detection~\cite{yenala2017convolutional}, 
A* search and Markov noisy channel models for online spell correction~\cite{duan2011online}, and character RNNs~\cite{wang2018realtime}. There have also been attempts to \textit{generate} effective QAC suggestions~\cite{maurya2023trie,maheswaran2024dac,maheswaran2024dqac,mandal2026chat,mandal2026chat}.

\subsection{\system{}-like systems}
Unfortunately, there has been no work on the \system{} problem which we study in this paper. There have been some attempts on type-ahead completions for Teams~\cite{trajanovski2021does} and email composition systems like GMail~\cite{chen2019gmail} and Outlook~\cite{trajanovski2021does}, which help users to compose, whereas \system{} helps the users in finding information (Navigation). Further, existing WebQAC methods cannot be trivially adapted for the \system{} task since it involves careful modeling of the additional document context. Our proposed system can be effectively integrated with in-document search systems like KTRL+F~\cite{oh2024ktrl+} to enhance their functionality. 

\subsection{Constrained Decoding}
Recent studies on constrained decoding have focused on constraining the output of language models, primarily using hard constraints. Early methods such as grid beam search \cite{hokamp2017lexically} and dynamic beam allocation \cite{post2018fast} introduced mechanisms to enforce the inclusion of specific words or phrases in the generated text. In parallel, grammar-based decoding approaches \cite{geng2023grammar} have emerged as another direction for structured generation.
However, such hard constraint mechanisms often limit the generative flexibility of models, making them less suitable for open-ended tasks such as WebQAC or \system{}, where a balance between adherence and fluency is crucial. Recently, prefix trie-based methods \cite{chan2025efficient} have been explored to improve beam search efficiency during decoding. 

Although, there exists work on utilizing tries in constrained generation in other tasks of information retrieval ~\cite{DeCao2021Autoregressive,Bevilacqua2022AutoregressiveSE}
,no prior work exists that applies an adaptive penalty schedule at decoding-time on trie constrained generation.In document-level QAC, such adaptive constraints become even more meaningful where the document context encoded in the trie naturally guides generation, allowing the model to generate plausible and context-aligned completions while maintaining the advantages of neural generation.

\section{\system{} Problem Formulation}

Let $\mathcal{V}$ be the vocabulary of all terms and $\mathcal{D}$ be a corpus of documents. Given a query prefix $p$ consisting of a sequence of tokens $(w_1, \dots, w_k)$, the goal of \textbf{Query Auto-Completion (QAC)} is to generate the optimal completion suffix $s$ such that the full query $q = p \oplus s$ maximizes the posterior probability.

\paragraph{Existing WebQAC (Global Optimization):}
Standard WebQAC operates in an ``open-world'' setting where the objective is to maximize the likelihood of $q$ given the prefix $p$ and the global user search history $\mathcal{H}_{global}$. Mathematically, this approximates the marginal probability over all possible latent contexts (or documents $d$):

\begin{equation}
    q^*_{web} = \operatorname*{argmax}_{q \in \mathcal{V}^*} P(q \mid p, \mathcal{H}_{global})
\end{equation}

Consequently, WebQAC systems are biased toward ``head'' queries that are frequent across the entire corpus, often ignoring specific document contexts. The support of the distribution is effectively the entire vocabulary $\mathcal{V}$, meaning any plausible string has non-zero probability ($P > 0$).

\paragraph{\system{} (Conditional Constrained Optimization):}
In contrast, \system{} is a ``closed-world'' task where the completion is strictly conditioned on a specific observed document $d_{curr}$ as well as the global user search history for this document $H_{curr}$. The objective changes to maximizing the conditional probability:

\begin{equation}
    q^*_{doc} = \operatorname*{argmax}_{q \in \mathcal{V}^*} P(q \mid p, d_{curr}, \mathcal{H}_{curr})
\end{equation}



Unlike WebQAC, where unseen queries are smoothed, \system{} treats suggestions outside the document's scope as hallucinations. This fundamental shift requires models to suppress the global token popularity and exclusively rely on the local likelihood $P(q \mid p, d_{curr})$, necessitating the trie-guided constrained decoding approach proposed in this work.

\section{\system{} Dataset}
\label{sec:dataset}
We utilize the ORCAS~\cite{craswell2020orcas} dataset which contains 1.4M documents and 10M distinct real-world user queries. We illustrate our dataset construction process in Fig.~\ref{fig:datapipeline} and describe it in this section.

\noindent\textbf{Obtaining Frequency Counts and Initial Preprocessing.} 
We perform the following preprocessing steps: (1) We retain queries that are at least 3 characters long. (2) We remove duplicate query-document (QD) pairs. (3) We retain documents with more than 10 queries but fewer than 500 queries to ensure a balanced representation. (4) We choose only those query-document pairs where the document exists in the TREC dataset\footnote{\url{https://msmarco.z22.web.core.windows.net/msmarcoranking/msmarco-docs.tsv.gz}\label{trecData}}. Thus, we obtain document text content from the TREC dataset.

Next, we create train/validation/test dataset splits while considering the temporal aspects of the dataset. To do this, we look up the (query, document) pairs from the ORCAS dataset against Bing logs 
from Jul-Aug 2022\footref{oldDataExplanation}, and prepare the training splits from a 30 day window, a validation split from a 4-day window, and a test set from a 10-day period. This process also helps us obtain the number of times a document was clicked for a given query which in turn is helpful in ranking suggestions for QAC.

\noindent\textbf{Query Augmentation via Similar Queries.} 
To augment the ORCAS dataset, for each unique clicked query $Q$, we start by collecting 100 most similar queries $Q'$  from Bing query logs (Jul-Aug 2022\footnote{We use an older time point to match timeline of original queries in ORCAS.\label{oldDataExplanation}}) using cosine similarity over DeBERTa-v3-base\footnote{\url{https://huggingface.co/microsoft/deberta-v3-base}} embeddings. 
To ensure the quality and relevance of these similar queries for the \system{} task, we applied a series of rigorous filtering steps. First, to create a publicly shareable dataset, we retained only those similar queries that appeared verbatim within the content of some document in the collection. 
Second, we remove any similar queries that were near-duplicates (Levenshtein distance $\leq$ 1) or if they were already part of clicked queries in ORCAS for the same document. 




\noindent\textbf{Relevance Labeling.}
A critical component of our dataset is the relevance label for each query-document pair. We use GPT-4\footnote{\url{https://cdn.openai.com/papers/gpt-4.pdf}\label{gpt4}} as a binary relevance classifier to evaluate whether each query-document pair is relevant or not, going beyond historical clicks. This helps in removing false positives due to user behavior patterns or popularity biases. Overall, this leads to significant enhancement to the dataset where 48.79\% are the original clicked queries $Q$ and the remaining are similar queries $Q'$.

\noindent\textbf{Click Popularity Estimation.}
As query and document click counts are not available for these similar queries (since they did not directly lead to clicks to the documents), we estimate them based on our original dataset as follows. For a similar query $Q'$, we identify its top-5 closest matches from the existing queries $Q$ in our dataset (which have frequency counts), and calculate a weighted average of their historical click counts, where the weight is the cosine similarity between the similar query $Q'$ and its top-5 most similar queries. A similar methodology was used to assign click counts to a similar query $Q'$ for a document by using counts of top-5 closest clicked queries of that document.

\noindent\textbf{Test Dataset Creation.}
Within the test set, we create 4 distinct test splits based on the presence of test set queries and documents in the training dataset. Specifically, if both the query and the document are present in both the training and test sets, we call this subset the ``seen query-seen document (\textbf{SS})'' test set. Conversely, if neither the query nor the document appears in the training set, the corresponding subset is labeled as the ``unseen query-unseen document (\textbf{UU})'' test set. Additionally, we define two other splits: the ``unseen query-seen document (\textbf{US})'' test set, where the query is absent from the training set but the document is present, and the ``seen query-unseen document (\textbf{SU})'' test set, where the query is seen during training but the document is not. Each split in the test set contains 3,000 (query, document) pairs. Table~\ref{tab:datasetStats} shows the statistics of various subsets of our dataset.

For each (query, document) pair in the train set, a sample is created in the dataset by randomly choosing a split point within the query. We refer to the string to the left (right) of the split point in the query as the prefix (suffix or completion). Thus, each sample consists of a prefix and a document as input and the goal is to generate the suffix.

\begin{table}[!t]
 \centering 
 \scriptsize
 \tabcolsep4pt
 \begin{tabular}{lccccc} 
 \toprule
 Dataset & Docs &  (Query, Doc) Pairs & Prefixes & Avg Query Len\\ 
 \midrule
 Train & 22,453 & 316,813 & Dynamic & 17.1 \\
 Validation  & 7,266 & 31,682 & Dynamic & 16.8 \\
 Test-Seen Q Seen D & 2,611  & 3,000 & 53,862 &19.0  \\
 Test-Seen Q Unseen D  & 712 & 3,000	& 52,678 & 18.5\\
 Test-UnSeen Q Seen D & 2,485  & 3,000 & 54,145 &20.5\\
 Test-Unseen Q UnSeen D & 1,068 & 3,000 & 52,488 & 17.1\\
 \bottomrule
 \end{tabular}
 \caption{Dataset Statistics for different Dataset Splits. Length is in characters. ``Dynamic'' implies that query in the (query, doc) pair is split into prefix and suffix by choosing a random split point in every batch.}
 \label{tab:datasetStats}
\end{table}

\section{Methods for \system{}}
\label{sec:approaches}
We follow various modeling strategies for \system{} that gives a complete view at the spectrum of trade-off between accuracy and latency. We investigate several ML and DL approaches for the task, including trie-based methods, QueryBlazer, and neural language models. 

\begin{figure}
    \centering
    \includegraphics[width=\linewidth]{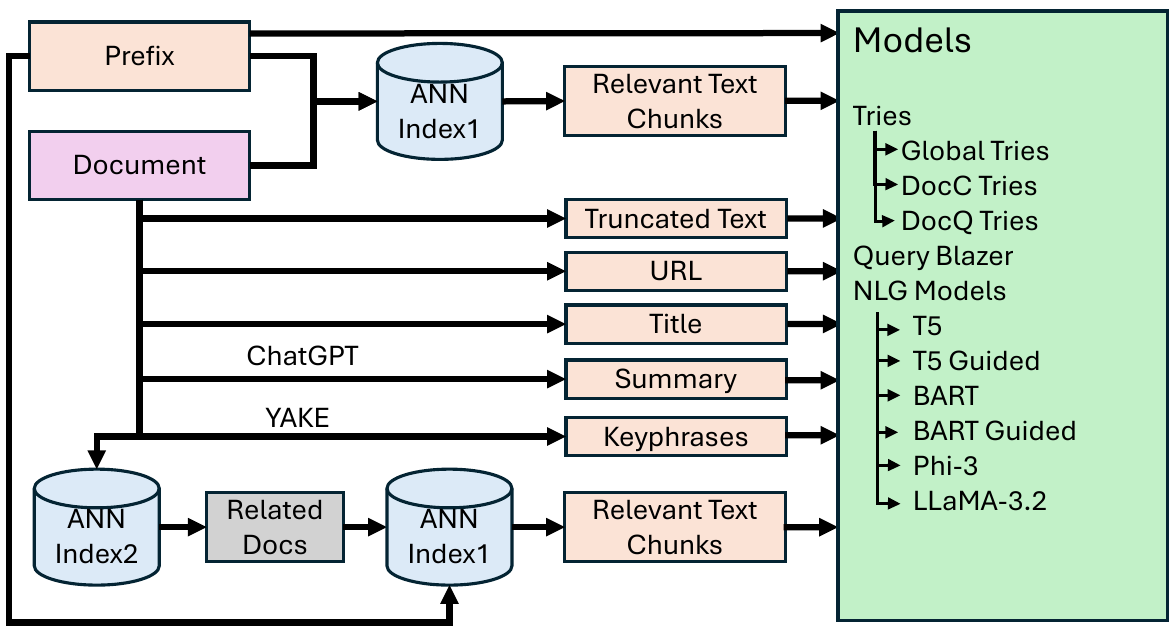}
    \caption{Input Representations and \system{} Methods}
    \label{fig:arch}
\end{figure}


\subsection{\system{} Tries}
To improve the coverage of main trie, we design an alternative method that utilizes a suffix-trie to handle prefixes with no matches in the main trie 
of the training queries. 
Specifically, we experiment with three different tries.

\begin{itemize}
    \item \textit{Global Query Trie}: All the queries of all the training documents are indexed into a single global trie. Given a test prefix and a document, the completions are selected using MPC from this global trie.
    
    \item \textit{Document Content (or DocC) Tries}: We build a DocC trie for each document by utilizing the ngrams of the document text. 
    \item \textit{Document Query (or DocQ) Tries}: We build a DocQ trie for each document using the document-specific subset of historical queries. Among our 4 test sets, DocQ tries cannot be made for test sets with unseen documents.   
\end{itemize}

\subsection{Neural Language Models}

We also experiment with popular Transformer-based models for generating \system{} suggestions. Specifically, we use the BART-base~\cite{lewis2019bart} and T5-small~\cite{roberts2019exploring} encoder-decoder models. Among Large Language models, we experiment with 2 models : Phi-3.5~\cite{abdin2024phi} and LLaMA-3.2~\cite{dubey2024llama} by LoRA-finetuning them on our training datasets. Particularly, we use google-t5/t5-small, facebook/bart-base, meta-llama/Llama-3.2-3B-Instruct and microsoft/phi-3.5-mini-instruct checkpoints. Refer Appendix~\ref{app:hyperparams} for hyperparameters.

The prefix and the completion constitute the source and target sequences, respectively, for these models. While training both the models, each training query is split stochastically into prefix and suffix. During inference, we use beam search to generate a ranked list of completions. 




\section{Trie-Guided Decoding}

Our analysis of performance of various methods reveals the following trade-off: trie-based methods excel at recall but fail at generalization, while generative language models excel at generalization but often lack relevance and precision.
In ``seen query, seen document (SS)'' scenarios, traditional trie-based approaches perform exceptionally well, as the task is primarily one of recall from a known set of queries. However, their rigidity becomes a critical failure point in ``unseen query, seen document (US)'' cases, where they are fundamentally unable to generate novel queries that are not present in their pre-compiled structure.
Conversely, LMs demonstrate strong performance on unseen query test sets by leveraging their generative capabilities to formulate novel, contextually relevant queries. However, their limitation is a lack of reliable grounding. In ``seen query, unseen document (SU)'' scenarios, an LM may fail to suggest a known, popular query, instead generating a fluent but less effective alternative. 

To address this, we propose an adaptive trie-guided mechanism that biases the model’s generation toward trie-conforming completions while retaining flexibility for contextual adaptation.
At each decoding step, we compute a bias term that is subtracted from the logits of tokens which do not match with the completions in the trie. This bias is annealed over time according to the length of the prefix and the diversity of the beam index, controlled by 3 hyperparameters: Initial bias, $\alpha$ and $\beta$. $\alpha$ controls decay with respect to prefix length. A higher $\alpha$ reduces the trie’s influence as prefix length increases. $\beta$ controls decay with respect to beam depth. 
\[
\text{annealed\_bias} = \text{initial\_bias} \cdot e^{-\alpha \cdot \text{length}} \cdot e^{-\beta \cdot \text{beam\_depth}}
\]
This encourages higher-ranked beams to follow the trie more strictly while allowing diversity in lower-ranked beams. Further, we also introduce an initial bias, which is a large initial penalty to strongly prioritize trie-conforming tokens at early decoding steps.



\begin{figure}
    \centering
    \includegraphics[width=1.00\columnwidth]{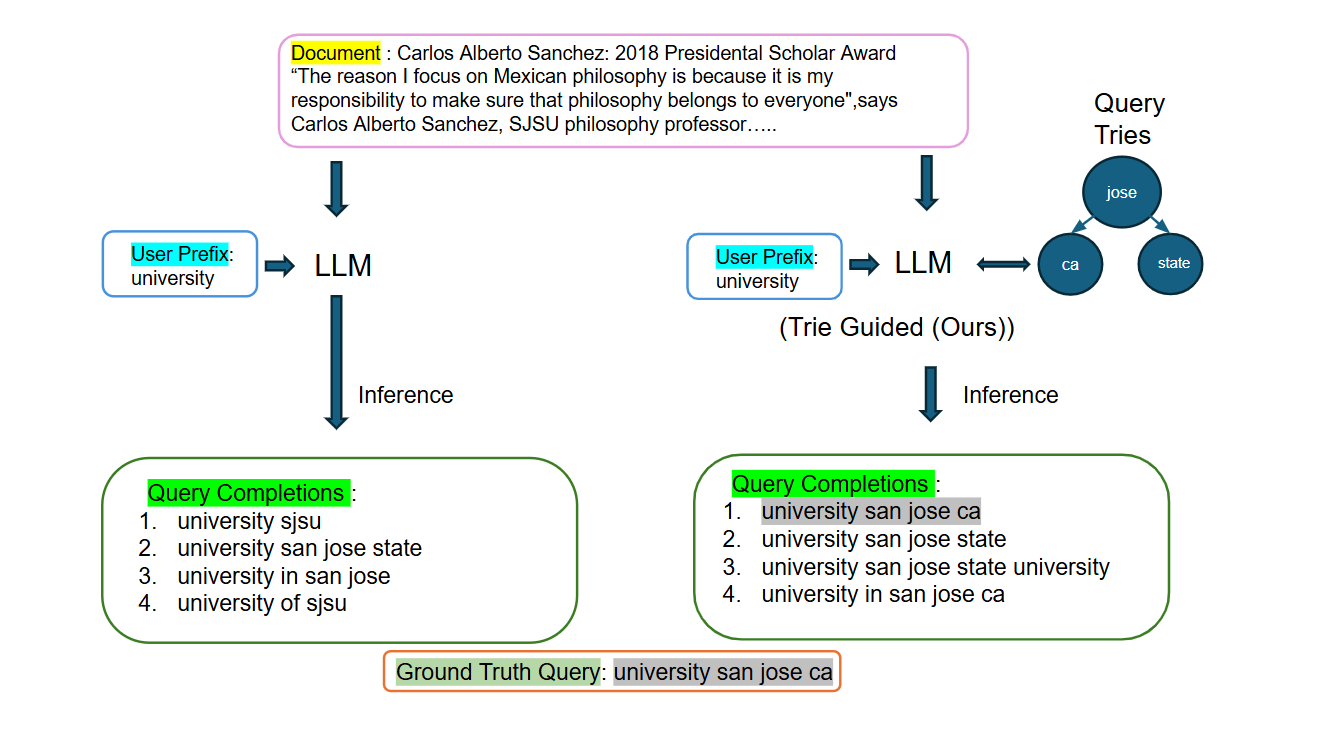}
    \caption{Illustration of Trie-Guided LLM vs Unguided LLM for a sample user prefix in \system{} setting.}
    \label{fig:intro}
\end{figure} 
Figure \ref{fig:intro} showcases the difference between a Trie Guided LLM vs an unguided LLM with an example. Our trie constrained system can validates if a query exists in a trie and updates its decoding process while inferencing, thereby helping the user in getting to the actual completion faster. The formal pseudocode for our constrained decoding strategy is presented in Algorithm \ref{alg:soft-trie-decoding}.
\begin{algorithm}[h]
\caption{Soft Trie-Guided Decoding}
\label{alg:soft-trie-decoding}
\DontPrintSemicolon
\KwIn{Language model $\mathcal{M}$,  
prefix $p$, trie $\mathcal{T}_r$, beam size $K$, 
initial bias $b_0$, decay parameters $\alpha, \beta$, 
maximum decoding steps $T$}
\KwOut{Top-$K$ query completions}
Initialize beam set $\mathcal{B}_0 \leftarrow \{p\}$

\For{$t = 1$ to $T$}{
    Initialize candidate set $\mathcal{C} \leftarrow \emptyset$
    
    \For{each beam $b_i \in \mathcal{B}_{t-1}$}{
        Compute logits $\mathbf{z}_t \leftarrow \mathcal{M}(b_i)$
        
        Retrieve valid next tokens         
        $\mathcal{V}_i \leftarrow \mathcal{T}_r.\textsc{ValidNextTokens}(b_i)$
        
        Compute annealed bias $\delta_i = b_0 \cdot e^{-\alpha \cdot |b_i|}
        \cdot e^{-\beta \cdot \text{rank}(b_i)}$
        
        \For{each token $v$ in vocabulary}{
            \If{$v \notin \mathcal{V}_i$}{
                $\mathbf{z}_t[v] \leftarrow \mathbf{z}_t[v] - \delta_i$
            }
        }
        
        $\mathbf{p}_t \leftarrow \textsc{Softmax}(\mathbf{z}_t)$
        
        Expand beam $b_i$ using top tokens from $\mathbf{p}_t$
        
        Add expanded beams to $\mathcal{C}$
    }
    
    Prune $\mathcal{C}$ to top-$K$ beams to form $\mathcal{B}_t$
}
Return $\mathcal{B}_T$
\end{algorithm}

\noindent\textbf{Tokenization Mismatch Problem}. 
A core challenge in applying trie-based guidance to language models for tasks like query suggestion is the mismatch between the granularity of user inputs, which is at character level, while the language model decodes at a subword level. Note that tries are built at a subword level to support such guidance. We build a serialized \texttt{BytesTrie} for fast lookup during decoding. 

Given a prefix of a seen query, we need to ensure that its tokenization should match with a path in the trie. The last few characters of a partial token should also match with a node in the trie. For example, for the query ``machine learning'', it is important to have ``machine lea'' also in the trie although tokenization of the original query just leads to two tokens ``machine'' and ``learning''. Hence, for each query \( q \) of the training dataset \(D\),  we generate all possible character-level prefix-suffix splits for all the queries, tokenize prefix and suffix separately, and store the overall tokenized string in the trie. However, at test time, this could lead to multiple redundant prefix path matches in the trie for super-strings. For example, consider a new query ``machine learning goals'' and prefix is ``machine learning''. This would match several paths in the trie including ``machine lea rning'', ``machine learn ing'' etc. To avoid a match with multiple paths in the trie, at trie creation time, we mark the transition from prefix to suffix by inserting a unique separator token (e.g., \texttt{[SEP\_SPLIT]}) between the last prefix and first suffix token IDs. At test time, we match the path corresponding to prefix+\texttt{[SEP\_SPLIT]} against the trie.

\section{Utilizing Document Context for \system{}}
\label{sec:docReps}
We evaluate our models discussed in Section~\ref{sec:approaches}, using different approaches to incorporate document context.  Specifically, we experiment with short input representations, longer document representations and retrieval augmented generation based approaches. For trie-based methods and QueryBlazer, we use the document content to rerank top 100 suggestions using cosine similarity based on sentence-BERT (\textit{msmarco-distilbert-base-v4}) embeddings. 

\subsection{Short Input}
To quantify the improvements obtained using the document context, we experiment with a \textbf{Prefix Only (P)} setting and \textbf{Title + URL + Prefix (P+TU)} setting - with document title and URL as input context.


\subsection{Longer Document Representations}
We experiment with three different longer document representations: document text, keyphrases (KPs) and summaries.
\begin{itemize}
    \item \textbf{Title + URL + Document + Prefix (P+TUD)}: We pass trimmed document content as input. Given our model's maximum context length of 512 tokens, we allocate a maximum of 32 tokens each for the title and URL and 352 tokens for document context and rest for the prefix and system prompt.
    \item \textbf{Title + URL  + KPs + Prefix (P+TUK)}: We use YAKE  ~\cite{campos2020yake} to extract keyphrases with a maximum n-gram length of three, and we limit the number of extracted phrases to fifty.
    \item \textbf{Title + URL  + Summary + Prefix (P+TUS)}: For this, we utilize ChatGPT 3.5 Turbo\footnote{https://chat.openai.com/chat} with 16,134 context window and generate offline summaries of up to 300 words and pass as representative document context 
\end{itemize}

\subsection{Retrieval Augmented Generation (RAG)}

\begin{itemize}
    \item \textbf{RAG using current document}:  In this setting, for a given query prefix, we utilize only the current document to extract relevant chunks. The relevant chunks are retrieved and ranked using sparse or dense similarity metrics as follows. 
    \begin{itemize}
    \item \textit{Sparse Retrieval (Sparse RAG)}.
    We retrieve top $k$ (=20) sentences from the documents with highest BM25 \cite{robertson2009probabilistic} similarity between the prefix and sentences.
    
    \item \textit{Dense Retrieval (Dense RAG)}.
    We split each document into fixed-size (200 characters) chunks with some overlap (30 characters) between two neighboring chunks. We index chunks using all-mpnet-base-v2~\cite{reimers2019sentence} embeddings and  FAISS~\cite{douze2024faiss}. Next, we use the vector similarity between the full prefix and  chunks to extract extract top $k$ (=20) chunks. 
    
    \end{itemize}

    \item \textbf{RAG using related documents (\textbf{Rel+Dense RAG})}:     Given the current document, we first obtain top 10 similar documents from the training set based on top similarity scores using \textit{msmarco-distilbert-base-v4} document embeddings and FAISS. \textit{msmarco-distilbert-base-v4} has been trained on the MS MARCO dataset, the same dataset from which our documents are derived. 
    
\end{itemize}


\begin{table}[!b]
    \centering
    \scriptsize
    \tabcolsep1.5pt
    \begin{tabular}{llllllllll}
    \toprule
    \textbf{Set} & \textbf{Model} & \textbf{Input} & \textbf{MRR} & \textbf{nDCG$_\alpha$} & \textbf{BLEU$_{rr}$} & \textbf{SBMRR} & \textbf{PPN} & \textbf{PRN} & \textbf{TES} \\
    \midrule
    SS & DocQ tries & P & \textbf{0.738} & \textbf{0.277} & \textbf{0.477} & \textbf{0.803} & \textbf{0.820} & 0.824 & 0.889 \\
    SS & DocQ tries & P + TUS & 0.688 & 0.274 & 0.472 & 0.778 & 0.814 & \textbf{0.826} & 0.889 \\
    SS & DocQ tries & Sparse RAG & 0.688 & 0.274 & 0.471 & 0.777 & 0.813 & \textbf{0.826} & 0.889 \\
    SS & DocQ tries & P + TUD & 0.686 & 0.273 & 0.471 & 0.776 & 0.813 & \textbf{0.826} & 0.889 \\
    SS & DocQ tries & Dense RAG & 0.686 & 0.273 & 0.471 & 0.776 & 0.813 & \textbf{0.826} & 0.889 \\
    SS & DocQ tries & Rel + Dense RAG & 0.686 & 0.273 & 0.471 & 0.775 & 0.813 & \textbf{0.826} & 0.889 \\
    SS & DocQ tries & P + TUK & 0.680 & 0.273 & 0.470 & 0.770 & 0.812 & \textbf{0.826} & 0.889 \\
    SS & DocQ-Guided BART & P + TUK & 0.720 & 0.080 & 0.349 & 0.793 & 0.692 & 0.790 & \textbf{0.896} \\
    SU & Global-Guided BART & P + TUK & \textbf{0.711} & 0.078 & 0.296 & 0.778 & 0.664 & 0.730 & 0.880 \\
    SU & LLaMA-3.2 & P + TUS & 0.462 & 0.088 & 0.277 & 0.668 & \textbf{0.740} & 0.691 & 0.813 \\
    SU & Global-Guided T5  & Sparse RAG & 0.706 & 0.076 & 0.299 & \textbf{0.779} & 0.674 & \textbf{0.737} & \textbf{0.882} \\
    SU & Global-Tries & P + TUS & 0.525 & \textbf{0.156} & \textbf{0.309} & 0.595 & 0.664 & 0.665 & 0.625 \\
    US & Phi-3.5 & P + TU & \textbf{0.401} & 0.052 & 0.282 & 0.557 & 0.710 & \textbf{0.670} & 0.756 \\
    US & LLaMA-3.2 & P + TUD & 0.381 & \textbf{0.077} & 0.283 & 0.548 & \textbf{0.724} & 0.658 & 0.730 \\
    US & Phi-3.5 & P + TUS & 0.392 & 0.052 & 0.280 & 0.554 & 0.714 & 0.667 & \textbf{0.760} \\
    US & LLaMA-3.2 & P + TU & 0.400 & 0.070 & \textbf{0.285} & \textbf{0.562} & 0.711 & 0.668 & 0.728 \\
    UU & Phi-3.5 & P + TU & \textbf{0.461} & 0.059 & 0.279 & 0.628 & 0.729 & 0.693 & 0.809 \\
    UU & LLaMA-3.2 & P + TUS & 0.442 & \textbf{0.085} & 0.280 & 0.615 & \textbf{0.743} & 0.680 & 0.792 \\
    UU & Phi-3.5 & Sparse RAG & 0.456 & 0.061 & 0.283 & \textbf{0.629} & 0.730 & \textbf{0.695} & 0.809 \\
    UU & Phi-3.5 & P + TUS & 0.452 & 0.060 & 0.278 & 0.625 & 0.736 & 0.693 & \textbf{0.821} \\
    UU & LLaMA-3.2 & Dense RAG & 0.452 & 0.084 & \textbf{0.285} & 0.624 & 0.731 & 0.686 & 0.785 \\
    UU & LLaMA-3.2 & P + TUK & 0.458 & 0.080 & 0.283 & \textbf{0.629} & 0.729 & 0.688 & 0.789 \\
    \bottomrule
    \end{tabular}
    \caption{Model and input combinations that result in at least one best metric value for any of the 4 test sets.}
    \label{tab:bestResults}
\end{table}

\begin{table}[h!]
\centering
\scriptsize
\begin{tabular}{llllllllll}
\toprule
\textbf{Set} & \textbf{Model} & \textbf{Input} & \textbf{MRR} & \textbf{nDCG$_\alpha$} & \textbf{BLEU$_{rr}$} & \textbf{SBMRR} & \textbf{PPN} & \textbf{PRN} & \textbf{TES} \\
\midrule
SS & BART & P + TUK & 0.486 & 0.059 & 0.286 & 0.649 & 0.672 & 0.707 & 0.750 \\
SU & BART & P + TUK & 0.440 & 0.054 & 0.259 & 0.615 & 0.663 & 0.703 & 0.697 \\
SU & T5   & Sparse RAG & 0.425 & 0.051 & 0.260 & 0.617 & 0.677 & 0.710 & 0.685 \\
\hdashline
SS & DocQ-Guided BART   & P + TUK & 0.720 & 0.080 & 0.349 & 0.793 & 0.692 & 0.790 & 0.896 \\
SU & Global-Guided BART & P + TUK & 0.711 & 0.078 & 0.296 & 0.778 & 0.664 & 0.730 & 0.880 \\
SU & Global-Guided T5    & Sparse RAG & 0.706 & 0.076 & 0.299 & 0.779 & 0.674 & 0.737 & 0.882 \\
\bottomrule
\end{tabular}
\caption{Comparison of unguided vs. Trie-guided decoding (Ours). Dashed line separates the two groups.}
\label{tab:compareconstvsunc}
\end{table}


\begin{table}[h!]
\centering
\scriptsize
\begin{tabular}{llllllllll}
\toprule
\textbf{Set} & \textbf{Model} & \textbf{Input} & \textbf{MRR} & \textbf{nDCG$_\alpha$} & \textbf{BLEU$_{rr}$} & \textbf{SBMRR} & \textbf{PPN} & \textbf{PRN} & \textbf{TES} \\
\midrule
SS & Phi-3.5     & P + TUK     & 0.450 & 0.058 & 0.293 & 0.643 & 0.724 & 0.696 & 0.829 \\
SS & LLaMA-3.2   & P + TUK     & 0.446 & 0.079 & 0.296 & 0.645 & 0.724 & 0.689 & 0.791 \\
SU & Phi-3.5     & P + TUK     & 0.476 & 0.061 & 0.277 & 0.676 & 0.724 & 0.706 & 0.842 \\
SU & LLaMA-3.2   & P + TUK     & 0.477 & 0.083 & 0.279 & 0.676 & 0.725 & 0.699 & 0.808 \\
SU & Phi-3.5     & Sparse RAG  & 0.467 & 0.085 & 0.278 & 0.667 & 0.729 & 0.697 & 0.808 \\
SU & LLaMA-3.2   & Sparse RAG  & 0.473 & 0.063 & 0.279 & 0.670 & 0.730 & 0.706 & 0.833 \\
\hdashline
SS & DocQ-Guided BART   & P + TUK     & 0.720 & 0.080 & 0.349 & 0.793 & 0.692 & 0.790 & 0.896 \\
SU & Global-Guided BART & P + TUK     & 0.711 & 0.078 & 0.296 & 0.778 & 0.664 & 0.730 & 0.880 \\
SU & Global-Guided T5    & Sparse RAG  & 0.706 & 0.076 & 0.299 & 0.779 & 0.674 & 0.737 & 0.882 \\
\bottomrule
\end{tabular}
\caption{Comparison of large models (Phi-3.5, LLaMA-3.2) with our Trie-Guided models (T5 and BART). Dashed line separates the two groups.}
\label{tab:compareconstvsuncllm}
\end{table}

\begin{table*}[!t]
\centering
\scriptsize
\tabcolsep4pt
\begin{tabular}{lccccccccccccccc}
\hline
\textbf{Input} & \multicolumn{1}{c}{\textbf{MRR}} & \multicolumn{1}{c}{\textbf{$\alpha$N}} & \multicolumn{1}{c}{\textbf{BLEU$_{RR}$}} & \multicolumn{1}{c}{\textbf{SBMRR}} & \multicolumn{1}{c}{\textbf{PPN}} & \multicolumn{1}{c}{\textbf{PRN}} & \textbf{TES} &  & \multicolumn{1}{c}{\textbf{MRR}} & \multicolumn{1}{c}{\textbf{$\alpha$N}} & \multicolumn{1}{c}{\textbf{BLEU$_{RR}$}} & \multicolumn{1}{c}{\textbf{SBMRR}} & \multicolumn{1}{c}{\textbf{PPN}} & \multicolumn{1}{c}{\textbf{PRN}} & \textbf{TES} \\ \cline{1-8} \cline{10-16} 
\multicolumn{1}{l}{} & \multicolumn{7}{c}{\textbf{SS, DocQ tries}} &  & \multicolumn{7}{c}{\textbf{US, LLaMA-3.2 }} \\ \cline{1-8} \cline{10-16} 
\textbf{P} & \multicolumn{1}{c}{\textbf{0.738}} & \multicolumn{1}{c}{\textbf{0.277}} & \multicolumn{1}{c}{\textbf{0.477}} & \multicolumn{1}{c}{\textbf{0.803}} & \multicolumn{1}{c}{\textbf{0.820}} & \multicolumn{1}{c}{0.824} &0.889 &  & \multicolumn{1}{c}{0.245} & \multicolumn{1}{c}{0.061} & \multicolumn{1}{c}{0.229} & \multicolumn{1}{c}{0.325} & \multicolumn{1}{c}{0.669} & \multicolumn{1}{c}{0.572} & 0.499 \\ 
\textbf{P+TU} & \multicolumn{1}{c}{0.687} & \multicolumn{1}{c}{0.274} & \multicolumn{1}{c}{0.470} & \multicolumn{1}{c}{0.776} & \multicolumn{1}{c}{0.815} & \multicolumn{1}{c}{0.825} & 0.889 &  & \multicolumn{1}{c}{0.400} & \multicolumn{1}{c}{0.070} & \multicolumn{1}{c}{\textbf{0.285}} & \multicolumn{1}{c}{\textbf{0.562}} & \multicolumn{1}{c}{0.711} & \multicolumn{1}{c}{0.668} & 0.728\\ 
\textbf{P+TUD} & \multicolumn{1}{c}{0.686} & \multicolumn{1}{c}{0.273} & \multicolumn{1}{c}{0.471} & \multicolumn{1}{c}{0.776} & \multicolumn{1}{c}{0.813} & \multicolumn{1}{c}{\textbf{0.826}} & 0.889 &  & \multicolumn{1}{c}{0.388} & \multicolumn{1}{c}{0.072} & \multicolumn{1}{c}{0.283} & \multicolumn{1}{c}{0.551} & \multicolumn{1}{c}{0.713} & \multicolumn{1}{c}{0.663} & 0.724 \\ 
\textbf{P+TUK} & \multicolumn{1}{c}{0.680} & \multicolumn{1}{c}{0.273} & \multicolumn{1}{c}{0.470} & \multicolumn{1}{c}{0.770} & \multicolumn{1}{c}{0.812} & \multicolumn{1}{c}{\textbf{0.826}} & 0.889 &  & \multicolumn{1}{c}{0.387} & \multicolumn{1}{c}{0.071} & \multicolumn{1}{c}{0.283} & \multicolumn{1}{c}{0.552} & \multicolumn{1}{c}{0.707} & \multicolumn{1}{c}{0.664} & 0.714 \\ 
\textbf{P+TUS} & \multicolumn{1}{c}{0.688} & \multicolumn{1}{c}{0.274} & \multicolumn{1}{c}{0.472} & \multicolumn{1}{c}{0.778} & \multicolumn{1}{c}{0.814} & \multicolumn{1}{c}{\textbf{0.826}}
& 0.889 &  & \multicolumn{1}{c}{0.381} & \multicolumn{1}{c}{\textbf{0.077}} & \multicolumn{1}{c}{0.283} & \multicolumn{1}{c}{0.548} & \multicolumn{1}{c}{\textbf{0.724}} & \multicolumn{1}{c}{0.658} & 0.730 \\ 
\textbf{Sparse RAG} & \multicolumn{1}{c}{0.688} & \multicolumn{1}{c}{0.274} & \multicolumn{1}{c}{0.471} & \multicolumn{1}{c}{0.777} & \multicolumn{1}{c}{0.813} & \multicolumn{1}{c}{\textbf{0.826}} & 0.889 &  & \multicolumn{1}{c}{0.388} & \multicolumn{1}{c}{0.074} & \multicolumn{1}{c}{0.284} & \multicolumn{1}{c}{0.550} & \multicolumn{1}{c}{0.713} & \multicolumn{1}{c}{0.662} & 0.716 \\ 
\textbf{Dense RAG} & \multicolumn{1}{c}{0.686} & \multicolumn{1}{c}{0.273} & \multicolumn{1}{c}{0.471} & \multicolumn{1}{c}{0.776} & \multicolumn{1}{c}{0.813} & \multicolumn{1}{c}{0.824} & 0.889 &  & \multicolumn{1}{c}{0.384} & \multicolumn{1}{c}{0.075} & \multicolumn{1}{c}{0.285} & \multicolumn{1}{c}{0.546} & \multicolumn{1}{c}{0.711} & \multicolumn{1}{c}{0.660} & 0.714 \\
\textbf{Rel+Dense RAG} & \multicolumn{1}{c}{0.686} & \multicolumn{1}{c}{0.273} & \multicolumn{1}{c}{0.471} & \multicolumn{1}{c}{0.775} & \multicolumn{1}{c}{0.813} & \multicolumn{1}{c}{\textbf{0.826}} & 0.889 &  & \multicolumn{1}{c}{0.370} & \multicolumn{1}{c}{0.067} & \multicolumn{1}{c}{0.281} & \multicolumn{1}{c}{0.543} & \multicolumn{1}{c}{0.696} & \multicolumn{1}{c}{0.659} & 0.697 \\ \cline{1-8} \cline{10-16} 
\multicolumn{1}{l}{} & \multicolumn{7}{c}{\textbf{SU, Global-Guided T5 }} &  & \multicolumn{7}{c}{\textbf{UU, LLaMA-3.2}} \\ \cline{1-8} \cline{10-16} 
\textbf{P} & \multicolumn{1}{c}{0.384} & \multicolumn{1}{c}{0.042} & \multicolumn{1}{c}{0.221} & \multicolumn{1}{c}{0.463} & \multicolumn{1}{c}{0.593} & \multicolumn{1}{c}{0.636} & 0.552 &  & \multicolumn{1}{c}{0.241} & \multicolumn{1}{c}{0.060} & \multicolumn{1}{c}{0.210} & \multicolumn{1}{c}{0.318} & \multicolumn{1}{c}{0.665} & \multicolumn{1}{c}{0.568} & 0.487 \\ 
\textbf{P+TU} & \multicolumn{1}{c}{0.538} & \multicolumn{1}{c}{0.059} & \multicolumn{1}{c}{0.267} & \multicolumn{1}{c}{0.675} & \multicolumn{1}{c}{0.662} & \multicolumn{1}{c}{0.714} & 0.771 &  & \multicolumn{1}{c}{0.460} & \multicolumn{1}{c}{0.078} & \multicolumn{1}{c}{0.282} & \multicolumn{1}{c}{0.626} & \multicolumn{1}{c}{0.730} & \multicolumn{1}{c}{0.688} & 0.789 \\ 
\textbf{P+TUD} & \multicolumn{1}{c}{0.552} & \multicolumn{1}{c}{0.060} & \multicolumn{1}{c}{0.270} & \multicolumn{1}{c}{0.690} & \multicolumn{1}{c}{0.662} & \multicolumn{1}{c}{0.718} & 0.788 &  & \multicolumn{1}{c}{0.449} & \multicolumn{1}{c}{0.080} & \multicolumn{1}{c}{0.281} & \multicolumn{1}{c}{0.619} & \multicolumn{1}{c}{0.734} & \multicolumn{1}{c}{0.685} & 0.792 \\ 
\textbf{P+TUK} & \multicolumn{1}{c}{0.549} & \multicolumn{1}{c}{0.060} & \multicolumn{1}{c}{0.268} & \multicolumn{1}{c}{0.686} & \multicolumn{1}{c}{0.664} & \multicolumn{1}{c}{0.715} & 0.787 &  & \multicolumn{1}{c}{0.458} & \multicolumn{1}{c}{0.080} & \multicolumn{1}{c}{0.283} & \multicolumn{1}{c}{\textbf{0.629}} & \multicolumn{1}{c}{0.729} & \multicolumn{1}{c}{0.688} & 0.789 \\ 
\textbf{P+TUS} & \multicolumn{1}{c}{0.528} & \multicolumn{1}{c}{0.058} & \multicolumn{1}{c}{0.264} & \multicolumn{1}{c}{0.672} & \multicolumn{1}{c}{0.669} & \multicolumn{1}{c}{0.711} & 0.805 &  & \multicolumn{1}{c}{0.442} & \multicolumn{1}{c}{\textbf{0.085}} & \multicolumn{1}{c}{0.280} & \multicolumn{1}{c}{0.615} & \multicolumn{1}{c}{\textbf{0.743}} & \multicolumn{1}{c}{0.680} & 0.792 \\ 
\textbf{Sparse RAG} & \multicolumn{1}{c}{0.708} & \multicolumn{1}{c}{0.076} & \multicolumn{1}{c}{0.299} & \multicolumn{1}{c}{\textbf{0.779}} & \multicolumn{1}{c}{0.674} & \multicolumn{1}{c}{\textbf{0.737}} & \textbf{0.882} &  & \multicolumn{1}{c}{0.453} & \multicolumn{1}{c}{0.082} & \multicolumn{1}{c}{0.282} & \multicolumn{1}{c}{0.624} & \multicolumn{1}{c}{0.732} & \multicolumn{1}{c}{0.686} & 0.792\\ 
\textbf{Dense RAG} & \multicolumn{1}{c}{0.554} & \multicolumn{1}{c}{0.061} & \multicolumn{1}{c}{0.270} & \multicolumn{1}{c}{0.695} & \multicolumn{1}{c}{0.667} & \multicolumn{1}{c}{0.718} & 0.807 &  & \multicolumn{1}{c}{0.452} & \multicolumn{1}{c}{0.084} & \multicolumn{1}{c}{\textbf{0.285}} & \multicolumn{1}{c}{0.624} & \multicolumn{1}{c}{0.731} & \multicolumn{1}{c}{0.686} & 0.785 \\ 
\textbf{Rel+Dense RAG} & \multicolumn{1}{c}{0.553} & \multicolumn{1}{c}{0.060} & \multicolumn{1}{c}{0.268} & \multicolumn{1}{c}{0.688} & \multicolumn{1}{c}{0.667} & \multicolumn{1}{c}{0.717} & 0.798 &  & \multicolumn{1}{c}{0.446} & \multicolumn{1}{c}{0.076} & \multicolumn{1}{c}{0.280} & \multicolumn{1}{c}{0.617} & \multicolumn{1}{c}{0.716} & \multicolumn{1}{c}{0.685} & 0.769 \\ \hline
\end{tabular}
\caption{Results with different input representations, for the best performing model for each of the 4 test sets.}
\label{tab:mainResultsAllSettings}
\end{table*}

\section{Evaluation Metrics }
We categorize our metrics for \system{} into 2 categories and evaluate top 10 suggestions for each prefix on these metrics.

\subsection{Primary Metrics} 
We prioritize the evaluation metrics  that directly quantify the success of the system in meeting the user's core objective: finding the correct document content with minimal physical effort.

\textbf{Typing Effort Saved (TES)}: In \system{} scenarios (e.g., technical manuals, legal briefs, medical records), users often search for long, complex, or hard-to-spell domain-specific terms.
The TES metric, inspired by~\cite{trajanovski2021does}, is computed as $\text{TES} = 1-\frac{\text{No. of typed characters}} {\text{Query Length}}$. Unlike ranking metrics which evaluate a static list, TES simulates the dynamic, end-to-end user interaction (typing $\to$ looking $\to$ selecting). It answers the most critical practical question: ``Did this system actually make the query creation faster for users?'' In our \system{} settings, saving keystrokes is the ultimate goal.

\textbf{Mean Reciprocal Rank (MRR)}: As \system{} is a navigation task, where the goal is to locate a specific string instantly, the rank of the first correct answer is paramount.
In auto-completion, users rarely scan beyond the top 1 or 2 results. MRR strictly penalizes any system that buries the correct answer lower in the list. 

\subsection{Secondary Metrics}
These metrics are valuable for understanding why a model succeeds or fails, and for diagnosing specific error types (e.g., hallucination vs. drift) and can be used as a set of secondary metrics.

\textbf{Semantic Match (SBMRR)}: SBMRR gives credit for understanding the user's intent, even if the exact string matching failed. This helps researchers understand if a model is ``smart but imprecise'' (high SBMRR, low MRR). Instead of lexical match, we find a semantic match between the reference query and its auto-completions. We use a transformer based model, Sentence-BERT \cite{reimers2019sentence} \textit{(all-MiniLM-L6-v2)} to compute both the query and suggestions' representations. We consider a match if the semantic similarity is $\geq$0.9.

\textbf{Partial Match Metrics (PPN and PRN)}: To diagnose failure modes beyond binary success, we employ Partial Precision (PPN) and Partial Recall (PRN) NDCG. PPN penalizes ``hallucinated suffixes'' (e.g., suggesting ``Data Lake'' instead of ``Data Base'') by measuring how much of the suggestion is valid. Conversely, PRN identifies ``truncation errors'' (e.g., stopping at ``Machine'' instead of ``Machine Learning'') by measuring how much of the target phrase was captured, ensuring models balance precision with completeness.

\textbf{Diversity and N-gram Overlap ($\alpha$-NDCG, BLEU$_{RR}$)}: Clarke et al.~\cite{clarke2008novelty} defined the $\alpha$NDCG metric for evaluating diversity. For example, in a coding document, the prefix ``pro'' might match ``process'', ``program'', and ``protect''. A bad model might fill the top 10 slots with just variations of one word (``process'', ``processing'', ``processed''). $\alpha$-NDCG penalizes this redundancy.
 BLEU$_{RR}$ acts as a ``soft'' MRR by weighting n-gram overlap by reciprocal rank. In \system{}, it rewards models for capturing document-specific vocabulary, which distinguishes useful ``near-misses'' from completely irrelevant hallucinations.
 
\section{Results}
\label{sec:results}
\subsection{Overall Results}

For each of the 4 test sets, Table~\ref{tab:bestResults} shows the model and input combinations that result in at least one best metric value. All results are reported on the 4 \system{} test sets, each containing approximately 52,000-53,000 prefix-query pairs (see Table~\ref{tab:datasetStats}). 

We observe that for SS set, DocQ tries perform the best for all metrics except TES. Our guided decoding approach, leveraging DocQ tries, achieves the highest TES with BART-Base, outperforming DocQ tries, LLaMA-3.2 (3B), and Phi-3.5 (3B). For SU set, the DocQ tries are not available. In this setting, our BART model with KPs as input and Global Tries for guided decoding, provides the best MRR. T5 model with Sparse RAG as input and Global Tries for guided decoding achieves the highest TES and PPN. For US and UU, LLaMA-3.2 and Phi-3.5 lead to the best aggregated results. 

\subsubsection{Best Guided vs Unguided counterparts}
Table~\ref{tab:compareconstvsunc} compares the 3 best performing guided decoding models (from Table~\ref{tab:bestResults}) with their unguided counterparts. Across all metrics, the guided methods consistently outperform the unguided ones. For example, for the BART model, applying guided decoding yields a 48.1\% improvement in MRR (from 0.486 to 0.720) and a 19.4\% increase in TES (from 0.750 to 0.896) for SS test set. 
We observe similar gains for SeenQ-SeenD for both BART and T5 models. For T5 with Sparse RAG as input, applying guided decoding leads to 66\% improvement in MRR alone (from 0.425 to 0.706)  and 27.7\% improvement in TES (from 0.685 to 0.882).


\subsubsection{Best Guided Fine Tuned (BART and T5) vs Instruction Fine Tuned (Phi-3.5 and LLaMA)}
Table~\ref{tab:compareconstvsuncllm} shows comparison between our three best-performing guided decoding models with large models with the same input. Notably, both the T5- and BART-based models, when enhanced with trie decoding, consistently and substantially outperform the instruction fine-tuned LLaMA-3.2, while being approximately 52.5× smaller (T5) and 23× smaller (BART).


\begin{table}[!b]
    \centering
    \tabcolsep1pt
    \scriptsize
    \begin{tabular}{l@{\hskip 2pt}cccccccccccc}
    \toprule
     & \textbf{Global} & \textbf{DocC} & \textbf{DocQ} & \textbf{QB} & \textbf{T5} & \multicolumn{2}{c}{\textbf{T5 Guided}}&\textbf{BART}&\multicolumn{2}{c}{\textbf{BART Guided}}& \textbf{Phi} & \textbf{LLaMA} \\
     \cmidrule(lr){7-8} \cmidrule(lr){10-11}
     & & & & & & \textbf{DocQ} & \textbf{Global} & & \textbf{DocQ} & \textbf{Global} & \textbf{3.5} & \textbf{3.2} \\
    \midrule
    P            & 8  & 10  & 11  & 0.2 & 66 & 73 & 118 & 63 & 73 & 190 & 494  & 905 \\
    P+TU         & 31 & 36  & 30  & 38  & 62 & 71 & 113 & 58 & 66 & 194 & 748  & 983 \\
    P+TUD        & 33 & 38  & 35  & 37  & 71 & 73 & 123 & 64 & 72 & 197 & 1246 & 1156 \\
    P+TUS        & 28 & 33  & 31  & 31  & 64 & 73 & 116 & 65 & 74 & 206 & 1331 & 1138 \\
    P+TUK        & 29 & 34  & 30  & 31  & 66 & 72 & 118 & 66   & 74   & 204 & 1153 & 1114 \\
    Sparse RAG   & 32 & 38  & 34  & 40  & 76 & 88 & 127 & 74 & 84   & 211 & 1162 & 1103 \\
    Dense RAG    & 71 & 84  & 63  & 77  & 99 & 107 & 150 & 97 & 111 & 239 & 1203 & 1105 \\
    Rel+Dense RAG& 277 & 325 & 231 & 290 & 226 & 240 & 287 & 228 & 240 & 366 & 1344 & 1134 \\
    \bottomrule
    \end{tabular}
    \caption{Latency (in ms) per sample. T5, BART, Phi-3.5 and LLaMA-3.2 latencies are on GPU. Others on CPU.}
    \label{tab:latency}
\end{table}

\begin{figure*}[!t]
    \centering
    \includegraphics[width=0.8\linewidth]{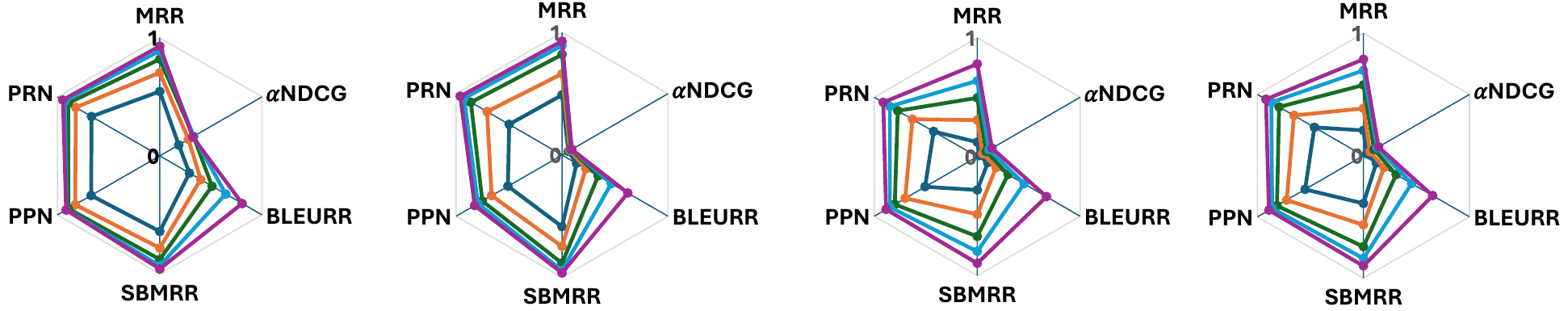}
    \caption{Performance Comparison across metrics for varying prefix lengths. Left to right: SS DocQ tries (P), SU Global-Guided T5  (Sparse RAG), US LLaMA-3.2 (P+TU), UU LLaMA-3.2 (P+TUK). Note TES cannot be computed for this experiment.}
    \label{fig:prefixLengths}
\end{figure*}

\begin{figure}
\begin{tcolorbox}[colback=blue!5!white, colframe=blue!80!black, title=]
\small
\texttt{<|im\_start|>system} \\
\texttt{[system](\#instructions)} \\
\texttt{\# Task} \\
Given a document, the following query was retrieved by an information retrieval (IR) system as a potential query a user might type for searching for content within the document. Your task is to accurately classify whether the query is truly relevant to the document or not. \\
\texttt{\# Input} \\
\texttt{Document: ``\{body\}''} \\
\texttt{Query: ``\{Query\}''} \\
\texttt{id : ``\{id\}''} \\
\texttt{docid : ``\{docid\}''} \\
\texttt{\# Output} \\
Provide your classification judgement of the query relevance for the document \emph{STRICTLY} in the following JSON format: \\
\texttt{\{} \\
\hspace*{3mm}\texttt{``query\_relevance'': bool,} \\
\hspace*{3mm}\texttt{``id'': string,} \\
\hspace*{3mm}\texttt{``Query'': string,} \\
\hspace*{3mm}\texttt{``docid'': string} \\
\texttt{\}} \\
\texttt{\# The query is relevant (true) if:} \\
-- The document contains specific information that directly answers the query. \\
-- The document provides background knowledge, explanations, or context that meaningfully relates to the query. \\
-- The document discusses entities, topics, or events explicitly mentioned in the query. \\
-- A user who issued this query would find the document useful or informative in addressing their information need. \\
\texttt{\# The query is not relevant (false) if:} \\
-- The document does not address the topic, entities, or intent expressed in the query. \\
-- The content is too vague, general, or off-topic to satisfy the query's information needs. \\
-- There is no logical or semantic connection between the query and the document content. \\
-- The document might mention some terms from the query, but in a completely unrelated context. \\
\texttt{<|im\_end|>}
\end{tcolorbox}
\caption{The system prompt used for document-query relevance classification using GPT4}
\label{fig:relevanceprompt}
\end{figure}

\begin{table*}[!t]
    \centering
    \scriptsize
    \begin{tabular}{|p{0.08\linewidth}|p{0.22\linewidth}|p{0.22\linewidth}|p{0.22\linewidth}|p{0.22\linewidth}|}
    \hline
&SS&SU&US&UU\\
\hline
Title & . & Francis I of France & \textbf{the third month of the year} & Walt Whitman \\
\hline
URL & \url{http://www.keybr.com/} & \url{https://en.wikipedia.org/wiki/Francis_I_of_France} & \url{http://www.timeanddate.com/calendar/months/} & \url{https://en.wikipedia.org/wiki/Walt_Whitman} \\
\hline
Query & \textbf{speed typing practice} & \textbf{king francis of france} & \textbf{the third month of the year} & \textbf{poems of walt whitman} \\
\hline
Prefix & spe & king & the thi & poems o \\
\hline
\hline
DocQ tries (P) & speed test practice, speed and eliminates, speed stayed, speed, speed test, speed for each, speed for every, speed for, speed stayed at & Trie cannot be created & Trie returned no matches due to log sparsity & Trie cannot be created \\
\hline
Phi-3.5 (P+TUK) & speed typing, speed of typing, speed in typing, speed to type, speed for typing, speed keybr, speeds of typing, speed keyboard, speed of keyboard typing, speed typ & king francis, king francis i, \textbf{king francis of france}, king francis the first, king of france, king francis 1, king francis iii, kings of france, king french, king francis 1st & the thirteen months, the thirty days, the thirteenth month, the thirteen month, the this month, the thirteen, the third month, the thirty one days & \textbf{poems of walt whitman}, poems of whitman, poems of walt, poems on walt whitman, poems of walter whitman, poems of walt whittman, poems on leaves of grass, poems of, poems of walt w, poems of walt watson \\
\hline
LLaMA-3.2 (P+TUK) & speed typing, speed type, speed of typing, speed typing test, \textbf{speed typing practice}, spell typing, spell typer, spell type, speed of type, spell keybr & king francis, king francis i, kings of france, king frances, king francisco, king francois, king frances i, kings francis, kings francis i, kings of franc & the third month, the thirteenth month, the this month, the this month is, the this months, the third month of, the thirteens, the thirtieth, the third & \textbf{poems of walt whitman}, poems of whitman, poems of walt whitan, poems of walt whiman, poems of walter whitman, \textbf{poems of walt whitman}s, poems of whitman's, poems of whitmans \\
\hline
DocQ-Guided BART (P+TUK) & \textbf{speed typing practice}, speed typing wpm, speed of your typing, speed of the keyboard, speed of a keyboard, speed of your keyboard, speed typing.com, speed of typing x, speed practice for keyboard, speed of a computer & Trie cannot be created & \textbf{the third month of the year}, the thir month september, the this month of the year, the thist month of the year, the things of the year, the thin months of the year, the this month, the third months of the year, the thir seasons of the year, the thingths of the year & Trie cannot be created \\
\hline
Global-Guided BART (P+TUK) & speed of typing, speed typing, speed of a computer, \textbf{speed typing practice}, speed checks, speak and type, speed website, speed of audio, speed up your computer, speed type game & kings of france, \textbf{king francis of france}, kings in france, king henry viii, king francis, king henry vii, king if england, kings, king of france timeline, kings uk & the third person, the things of the year, \textbf{the third month of the year}, the thingths of the year, the third months of the year, the thing months of the year, the thir month september, the things in the year, the thirds of the year, the this month of the year & \textbf{poems of walt whitman}, \textbf{poems of walt whitman} books, poems of walt walt breath, poems of walt walt Whitman, poems of walt walt, poems of walt rhodea, poems of walt rhodeo, poems of walt walt rock, poems of walt chattman, poems of walt walt cod \\
\hline
BART (P+TUK) & speed your typing, speed of typing, speed to type, speed of your typing, speed typing, speed writers, speed of the keyboard, speed of a keyboard, speed of your keyboard, speed your keyboard & kings of france, king of france, king francis viii, king and queen of france, king francis i, \textbf{king francis of france}, king's death of france, king's history of france, king francis the great, kings of france list & \textbf{the third month of the year}, the thir month september, the this month of the year, the thist month of the year, the things of the year, the thin months of the year, the this month, the third months of the year, the thir seasons of the year, the thingths of the year & \textbf{poems of walt whitman}, poems of walt walt, poems of walt, poems of walt codman, poems of walt norfolk, poems of walt washington, poems of walt potman, \textbf{poems of walt whitman} books, poems of walt walt breath, poems of walt marshall \\
\hline
    \end{tabular}
    \caption{Sample predictions from our best models comparing guided and unguided models.}
    \label{tab:caseStudies}
\end{table*}

\textbf{\subsection{Ablation Studies}}
\subsubsection{Analysis of input representations}
We present impact of each input representation  in Table~\ref{tab:mainResultsAllSettings}. \textbf{For SS set}, the best models are with DocQ tries with various inputs followed by our guided decoding models BART and then T5 with various inputs. \textbf{For SU set}, our global trie-guided T5 model with Sparse RAG has the highest TES, and global trie-guided BART with keyphrases as input has the best MRR. Notably, these models outperform global tries by a large margin. 
\textbf{For US set}, LLaMA-3.2 and Phi-3.5 with title and URL perform the best in all metrics except PRN. \textbf{For UU set}, LLaMA-3.2 and Phi-3.5, with either the title and URL, or both title and URL along with a document or summary as context, achieve the best results. Also, for both \textbf{US and UU set}, we observe the next best models to be BART and T5; tries perform the worst. Thus, for unseen queries, leveraging neural LMs is recommended.


\subsection{Other Analysis}
\subsubsection{Varying Prefix Length Analysis}
We assess performance of \system{} models in these prefix length categories: 1-5, 6-10, 11-15, 16-20 and 20+ characters. Fig.~\ref{fig:prefixLengths} shows results for our best models across all the 4 test sets. We observe that for each set across all metrics, results improve as the prefix length increases. We believe this is because longer prefixes provide the best clues for the models to predict accurate suggestions.




\subsubsection{Qualitative Analysis}


Table~\ref{tab:caseStudies} shows predictions for 4 samples (one from each test set) across all models, for best guided models and their unguided counterparts. We observe that DocQ tries cannot predict anything for 3 of these samples due to query log sparsity for those documents. 
For the SS sample, we observe that all guided models are able to correctly show query suggestion matching the user input query while their unguided counterparts are unable to do so for the prefix ``spe'' for the given  query ``speed typing practice''. In the SU scenario, the guided decoding models of BART are able to complete the prefix ``king'' by leveraging Global Tries. In the UU case, we observe that providing additional context like Sparse RAG to T5 and keyphrases to BART helped with getting the correct suggestion.

\subsubsection{Latency Analysis}

Table~\ref{tab:latency} reports latency across various methods computed using batch size of 1. Latency for neural LMs is on GPUs while for other models, we report latency on CPUs. Clearly, neural models have high latency compared to tries or QB.   

Our guided decoding models are highly practical for real-world deployment too. They incur only a modest 12 ms latency increase for T5 Guided DocQ  + Sparse RAG (from 76 ms to 88 ms) for unguided vs guided scenarios, while remaining $\sim$15x faster than both LLaMA-3.2 3B and Phi-3.5 3B. Similarly, DocQ-Guided BART + (P+TUK) model incurs a latency of increase of just 8 ms (from 66 to 74 ms) while remaining 15 times faster than LLaMA-3.2 3B and Phi-3.5 3B. Lastly, Global-Guided BART + (P+TUK) model adds a latency of 138 ms on top of unguided version, but still being 5.5 times faster than LLaMA3 and Phi3. Although these models have not been optimized using any of the popular optimization methods like FasterTransformers or TensorRT-LLM, their latencies are already low enough to be deployed in practical settings. These results demonstrate that developing specialized decoding algorithms is a highly effective and resource-efficient alternative to simply scaling up model size.

\section{Conclusion}
In this paper, we propose a novel task of \system{} for in-document query auto completions.
We also release the newly created \system{} benchmark dataset. We conduct extensive experiments by integrating various strategies, including RAG, using document summaries as context for LLMs, and combining these approaches with traditional methods such as tries. We establish strong baselines for these techniques and further evaluate larger language models like LLaMA and Phi-3. Building on this, we introduce a novel adaptive trie-guided decoding framework that enhances the performance of  language models like T5 and BART by guiding their outputs rather than imposing hard constraints. Our results demonstrate that this approach significantly outperforms the established baselines and even surpasses large instruction-tuned models. Overall, our findings showcase the potential of adaptive trie-guided decoding as a practical and efficient method for controlled language generation, and encourage further research in context-aware and nuanced text generation.

\appendix

\section{Query Document Relevance Classifier}
\label{app:query-doc-rel}
We utilize GPT4 with the system prompt as shown in Fig.~\ref{fig:relevanceprompt}  
to classify whether the query is relevant for a document or not.


\section{Hyper-parameter Settings}
\label{app:hyperparams}
\noindent\textbf{Compute}: Experiments were performed on a machine with 8 NVIDIA V100 GPUs for training neural models. For CPU-based experiments, we used an AMD Ryzen 9 7900X3D 12-Core Processor (4.40 GHz) with 64GB RAM.


\noindent\textbf{T5 and BART}: 
For T5, we use batch size=24,learning rate=1e-4,epochs=30,maximum input length=512, AdamW. For inference, we set beam size=25,max output length=48. For BART,we use 8 for document content (RAG, Summary, and KPs) and 18 for other experiments.

\noindent\textbf{Phi-3.5 and Llama-3.2}: Both models are fine-tuned using LoRA ~\cite{hu2021lora} for all linear layers (lora\_r=16, lora\_alpha=32,lora\_dropout=0.05), with AdamW,learning rate=5e-5,fp16 precision, max len=512,batch size=4 and epochs=5. Inference uses beam search with beam size=10, length penalty=1.0, and a 40-token output limit.




\noindent\textbf{Trie-Guided Decoding}: We perform extensive hyperparameter tuning across $\alpha \in \{0.05, 0.1, 0.2, 0.5\}$, $\beta \in \{0.05, 0.1, 0.2, 0.5\}$, $\text{bias} \in \{20, 30, 40\}$. For DocQ tries based guided decoding, we use (P+Sparse RAG), while for global-tries based guided decoding, we use (P+TUD) and run for SS and SU settings as they are the best performing models.
With global tries, the best performance on both SS and SU is obtained with $\alpha=0.1$, $\beta=0.05$, and $\text{bias}=40$. In contrast, for DocQ trie-guided decoding, the optimal configuration is $\alpha=0.1$, $\beta=0.2$, and $\text{bias}=40$ on SS, while $\alpha=0.5$, $\beta=0.2$, and $\text{bias}=20$ performs best on SU.
\bibliographystyle{ACM-Reference-Format}
\bibliography{refShort}


\end{document}